# The impact of three dimensional MHD instabilities on the generation of warm dense matter using a MA-class linear transformer driver.

P.-A. Gourdain, C. E. Seyler

Abstract

Warm dense matter is difficult to generate since it corresponds to a state of matter which pressure is order of magnitude larger than can be handled by natural materials. A diamond anvil can be used to pressurize matter up to one Gbar, this matter is at high density but at room temperature. High power lasers and heavy ion beams can generate warm dense matter but they cannot confine it long enough to allow measurements of quasi-static transport coefficients such as viscosity or heat conduction. We present here a third method to generate warm dense matter. It uses a pulsed-power driver which current rise time is substantially shortened by using a plasma opening switch, limiting the development of electrothermal instabilities. The switch relies on the implosion of a gas puff Z-pinch which carries most of the discharge current until the pinch reaches the sample. After that, the sample is compressed until it reaches the warm dense matter regime. Three-dimensional magnetohydrodynamics computations show that if the density of the gas is low enough no detectable instabilities (e.g. kinks and sausages modes) impede the remainder of the implosion.

## Introduction

Warm dense matter[1,2,3,4] (WDM) can be found at the center of gas giants[5,6] and Super-Earths[7]. Our ability to predict the size of a planet as a function of its mass hinges on the equations of state of WDM in the core and mantle of the planet. Yet this state of matter has elusive properties. The highly packed atoms form a strongly coupled lattice, which transport properties, like viscosity or heat conduction, cannot be inferred from plasma physics theories. The electrons are strongly degenerate. Yet their temperatures are too high to be able to derive their properties from density functional theory or condensed matter physics. While novel numerical methods[8] seem promising, experimental work is required to validate them. WDM is not difficult to produce using present-day facilities. Pulsed-power generators, high power lasers and heavy ion beams succeeded in producing materials samples of WDM, on the order of 1 to 10 times solid density at 1 eV. However, pulsed-power generators use an electrical current to heat the material sample (via Joule heating) and compresses the sample (via JxB forces). This current typically triggers electro-thermal instabilities which are detrimental to the quality of the material sample. This is because electro-thermal instabilities create hot spots which produce a mixture of WDM, solid, liquid and gas. These uncontrolled inhomogeneities are detrimental to line average diagnostics which cannot differentiate between the different states of matter. Ultimately the mixture is too heterogeneous to infer viable information from such diagnostics. Second, the plasma created by these instabilities expands rapidly and as its density decreases, heats up to the point where the plasma beta is on the order of unity. The magnetic compression scheme becomes relatively ineffective, trying to compress a hot plasma rather than a cold material sample.

However, using pulsed-power to generate warm dense matter has three main advantages. First, pulsed-power is a low cost technology which can pack copious amounts of energy into a material sample in a relatively short amount of time. On one hand, high-power lasers which are more than capable at moving electrons have proved very ill-suited at moving sizable amounts of the bulk material. While they can compress matter indirectly, via plasma ablation, they are limited in the mount of mass they can actually displace. On the other hand, pulsed-power drivers excel at moving mass, a necessary requirement in producing WDM. Second, pulsed-power drivers generate magnetic fields which can confine WDM for long periods of time (~10 ns). Heavy ion beams and lasers rely only on inertia to keep WDM together, which is effective only in the sub-ns time scale. Finally pulsed-power produce colder plasmas, in which the thermal energy is low enough to be comparable to the Fermi energy, hence the high level of degeneracy. Ultimately, different regimes can be explored with all three drivers. Yet lasers and beams have been heavily used due to the previously listed shortcomings of pulsed-power systems.

Previous work[9] showed that it may be possible to produce relatively homogeneous samples of WDM matter with pulsed-power drivers if a plasma switch is used in parallel with central rod, which will be turned into warn dense matter later in the pulse. Before the current flows, gas is injected between the anode and the cathode of the driver and pre-ionized. At the beginning of the current discharge, most of the current flows inside the plasma which also acts as a Faraday shield to the rod. As the current rises, the plasma collapses on axis and the current switches to the rod, starting to compress it. This current switch is 10 times faster than the current rise itself. In principle, the physical current switch would indicate that the system is a plasma opening switch. However, the collapse of the plasma on axis allows the current to switch. In this regard it is also a plasma closing switch. Regardless of its nature, the actual plasma motion is under a sleuth of axi-symmetric MHD instabilities, e.g. magnetic Rayleigh-Taylor, sausage. Since the simulations were two dimensional simulations, concerns were raised about three-dimensional instabilities. The present work explores their impact using PERSEUS[10]. The code is not capable of accounting for any types of micro-instabilities, limiting the scope of our results to macroscopic instabilities like kinks or sausage modes. Also this code does not have any quantum mechanical models. PERSEUS is a two fluid MHD code which advances electron motion implicitly using the electron inertia term inside the generalized Ohm's law, given in Eq. (1).

$$\frac{\partial \mathbf{J}}{\partial t} + \nabla \cdot \left( \mathbf{uJ} + \mathbf{Ju} - \frac{1}{en}\mathbf{JJ} \right) = \frac{1}{m_e}\nabla p_e + \frac{e^2 n}{m_e}\left( \mathbf{E} + \mathbf{u} \times \mathbf{B} - \frac{1}{en}\mathbf{J} \times \mathbf{B} - \eta \mathbf{J} \right) \quad (1)$$

While electron physics many not have a large impact in this study, the code runs as fast as a standard MHD code and leaving the electron physics present is not reducing code performances. It is important to use the generalized Ohm's law since the current switching scheme is strongly dependent of plasma resistivity. By using a simplified version of the Lee-More[11]-Desjarlais resistivity[12] $\eta$ for the high density plasma and the electron inertia in the low density plasma the proposed current switching scheme does not a use a "vacuum resistivity"[13], which is unphysical. We suppose the driver was a linear transformer driver[14] (LTD) with a current waveform following Eq. (2)

$$I(t) = I_{max} e^{-t/t_d} \sin\left(\frac{\pi t}{2 t_r}\right) \quad (2)$$

where $I_{max}$=1.2 MA, $t_r$=250ns is the rise time of the current and $t_d$=6$t_r$ is the damping time of the oscillations. The computational domain is 10mm long by 10mm wide and the height is 7mm. We used a total of 300x300x220 grid cells, giving a geometrical resolution of 50 µm. The anode is at the top and the cathode at the bottom. We use Al as our material rod (including electrode material) and He as our gas-prefill. Since Perseus has only one ion species, we artificially reduced the mass density of the gas to account for the increase in mass of the ion species. The initial gas density varies from $10^{16}$ to $4\times10^{16}$ cm$^{-3}$.

After this introduction, the paper looks at the efficiency of the current switching scheme, then focuses on the impact of gas density and rod radius on the properties of WDM. The best case scenario is explored is greater details using a partial domain decomposition to study at really high resolution a portion of the material rod. Our conclusions highlight that no major MHD instabilities develop across the whole rod in a manner significant enough to invalidate our approach.

## Current switching in three dimensions

Gas puffs have been used in the past as a mean to shock compress matter in Z-pinch configurations[15]. However, its efficiency as a plasma switch was never explored until recently[9]. That work used two dimensional simulations with constant background plasma density. Despite a very simplified model, the principle was demonstrated. However, its effectiveness in the presence of three-dimensional instabilities and with a realistic gas distribution needed to be explored. In this work we used a three-dimensional version of PERSEUS with a Gaussian gas puff profile given by:

$$n_{gas}(r, z) = n_0 \exp\left(\left[\frac{r - r_0}{2\sigma\omega(z)}\right]^2\right) / \omega(z) \qquad (3)$$

where r is the radius and z is the height. $\sigma\omega$ is the full-width half-maximum. $\omega(z)$ corresponds to the radial spread of the gas puff. This spread increases linearly as a function of the height z. We divided the gas density by $\omega(z)$ so that the total gas mass at a given height z is the same at all heights (i.e. mass conservation). The initial temperature of the gas was set to 0.25 eV. Figure 1-a shows the mass distribution for the gas puff 20 ns into the current discharge for an initial gas density of $16 \times 10^{22}$ m$^{-3}$. Despite the increase in density at the edge of the gas puff caused by the snowplow implosion, this figure shows the initial density of the gas though most of the gas puff volume. Random perturbations of up to 20% were added throughout the whole volume. These perturbations were not taken into account in the mass conserved Eq. (3).

Figure 1, together with Figure 2, shows the time evolution of the current as the gas puff implodes onto the axis. Figure 1-a plots the current density and the total ion number density in the three-dimensional computational box, 20 ns into the current discharge. At this time, the current density inside the rod is similar to the current density at the edge of the gas puff ($\sim 10^{10}$A/m$^2$). There is a physical reason why there is some current flowing in the rod this early in time. At the beginning of the discharge, where the plasma cylinder born from the gas puff is cold, the current flows both inside the gas and inside the rod. Rapidly, the plasma cylinder surrounding the rod heats up and prevents further current leaks to the rod. However, the initial current flowing inside the rod is trapped there for the rest of the discharge. In fact, it increases as the gas puff implodes, due to magnetic flux compression. At this point it would seem that the current switching scheme has failed since there is a non-negligible current density inside the rod.

To evaluate the amount of current flowing inside the rod we integrated the vertical current density crossing the domain mid-plane (z=0) inside a radius of 1mm (slightly bigger than the rod radius which is 0.8 mm). This current is labelled "rod current" in Figure 2. We have also integrated the total current flowing across the whole mid-plane. This current is labelled "total current" in Figure 2. At this point in the paper, we want to compare the current density shown in Figure 1, with the total current and the rod current, both shown in Figure 2. Figure 1 shows the spatial evolution of the current density with an initial gas puff density of $16 \times 10^{22}$ m$^{-3}$ which we labeled as "high density" in Figure 2. The total discharge current is 160kA 20 ns into the current rise. It is only 3 kA inside the rod. At this time the plasma cylinder surrounding the rod is hot enough to shield the rod from further current penetration. As evidence of this, we see that the current density throughout the rest of the plasma cylinder is much smaller ($\sim 10^8 A/m^2$). Since the plasma is carrying 98% of the current at this time, we see that the current switch is extremely efficient. Compared to the previous work[9], which was done with a current of 500 kA and a current rise time of 100 ns and where the gas was carrying 90 % of the current, we see remarkable improvement. We can also compare this case to the case where no initial gas puff is present, labelled "no gas" in Figure 2. At this time, we see that the whole current flows inside the rod.

At 40 ns into the discharge, in Figure 1-b, the current has mostly penetrated the whole plasma surrounding the rod. However, the rod current is only 4 kA for a total current of 330 kA. The plasma now carries almost 99 % of the total current. Figure 1-b also shows that part of the current flowing inside the anode (top cone) is redirected into the plasma and flows on the inner side of the plasma cylinder. The case with no gas present an interesting behavior. At 40 ns after the start of the current, there is only 289 kA of the current inside the rod. The rest of the current flows outside the rod. This is made possible by the plasma ablated from the rod. This plasma, heated by the current, expands away from the rod surface, carrying some current with it (41 kA). This ablation phase mixes solid with liquid and gas and it represents exactly what we are trying to avoid with the proposed setup where the initial gas puff strongly inhibits the ablation of the rod. While Figure 1-b does not show any sign of plasma instabilities we do expect magnetic Rayleigh-Taylor (MRT) instabilities to grow rapidly.

These instabilities clearly become visible in Figure 1-c, 60 ns into the current discharge, strongly disrupting the outer current path. At this point the current starts to switch to the rod. The total current is 473 kA and the rod current is 11.5 kA. They trigger the current switch. By 80 ns, the current flows mainly inside the rod. 614 kA of current flows inside while the rest of the current (37 kA) flows in the plasma left-over from the gas puff implosion. Using this system, we have achieved a current rise time of 20 ns. Figure 2 shows that after this time there is little difference on how the current flows with or without initial gas puff. In both cases the maximum compression happens around 160 ns. After the rod explodes, its diameter becomes larger than 1mm. At this point, the area in which the currents flow become larger than the area circumscribed by a 1-mm radius, accounting for the fast drop in current in Figure 2. As shown in previous work, gas puffs can be used efficiently as co-axial plasma opening switches. Typically, MRT instabilities start opening the switch. However, the switch fully open when the outer plasma as completely collapsed onto the rod. The two-dimensional analysis did not clearly show that MRT instabilities initiated the current switch.

The plasma switch is not perfect and some residual current flows inside the rod. If the rod is small enough, this current could generate rod breakdown and plasma production which we are trying to avoid to keep the rod as pristine as possible until compression starts. If we suppose that plasma formation is strictly provided by Ohmic heating, the resistivity used in MHD codes should be precise enough and the resolution high enough to capture when surface breakdown happens. According to our resistivity model, Figure 2 shows that we get plasma formation above 0.5 MG (called ablation threshold field thereafter), when the rod current departs from the total discharge current. Experimental measurements for Al have shown that this field should be close to 2 MG[16]. There is a disagreement between code and experiment which can be attributed to a resistivity model which cannot account for the sleuth of physical processes usually present in experiments (e.g. material anisotropy). However, this difference is reasonable enough that it does not contradict the argument this paper is trying to make. When the gas is initially present, the ablation threshold field is reached when the current time derivative is 5 times larger than the current derivative without gas prefill (see Figure 3). This large current derivative is required to have homogeneous current initiation[17], which will happen when the field reaches 0.5 MG.

## The impact of gas density on the generation of warm dense matter

Once the current switches fully to the rod, the compression begins. The major source of current inside the rod is coming from the collapsing plasma cylinder (which is a plasma sheet plagued with MRT instabilities by this time). Figure 3 shows that the current rise inside the rod, before the plasma sheet reaches it, is one order of magnitude smaller compared to the driver's. The current rise inside the rod becomes noticeable when the MRT instabilities start to tear apart in the plasma sheet. Most of the current transported by the sheet now flows inside the rod (t~80ns). However further increase of the total current driven by the pulsed-power generator does not flow solely inside the rod. A large portion of the subsequent current flows inside the low density plasma that surrounds the rod. It is interesting to note that this behavior is not caused by the initial gas prefill. Figure 2 shows not much difference in rod current with or without gas prefill. It demonstrates that the refill not only enhance the current rise in the rod but does not perturb much of the subsequent current paths compared to the case where there is no gas prefill.

However, MRT instabilities inside the gas have been carried over to the rod. Figure 4 shows that instabilities have created zones of inhomogeneities around the rod. This is unfortunate since the current switching scheme was designed to eliminate sources of instabilities at early times. Removing these initial seeds, plasma instabilities would be much difficult to grow. Yet, the same gas prefill, which increased the initial dI/dt, has really interfered with the compression of the rod at later times. Figure 4 shows that density non-uniformities are also enhanced by current non uniformities. They both trigger a sausage mode across the whole column. Non-uniform current densities lead to non-uniform heating, producing a warm dense matter sample with density *and* temperature variations along its length. Figure 5 shows a plot of ion number density versus temperature at peak compression on the linear scale. In this plot, the color scale corresponds to the kinetic pressure P. We computed the temperature (in eV) from the pressure given by the code using

$$\mathrm{T} = \frac{\mathrm{P}}{\mathrm{n}(1+\mathrm{Z})\mathrm{e}}. \tag{4}$$

This formula may not be appropriate to compute the actual temperature in this warm dense matter regime since the equation of state not fully known. Each datum corresponds to a cell value. We limited the sampling to a cylinder of 4 mm in height (2mm above and 2mm below the mid-plane) and 2 mm in radius. The spread in density and temperature is relatively large. However, it shows that we can reach densities on the order of $10^{30}$ m$^{-3}$ and temperature of 1 eV in a volume of several cubic millimeters.

Since MRT instabilities are the major source of non-uniformities of the warm dense matter sample, their reduction should help regaining homogeneity. We executed another set of numerical simulations. This time the initial gas density was set to $8\mathrm{x}10^{22}$ m$^{-3}$. Figure 6 shows the time evolution of the rod current for all three cases: $16\mathrm{x}10^{22}$ m$^{-3}$, $8\mathrm{x}10^{22}$ m$^{-3}$ and 0. We notice that the lower gas density still performs as well as a plasma switch, albeit switching the current 10 ns earlier compared to the case with higher current density. After the switch has occurred, the current evolution inside the rod is similar to the case with no gas prefill, showing that the prefill does not impact noticeably the current path compared to the no gas case. Figure 7 shows that all signs of instabilities are virtually gone when the rod is fully compressed. Both current and density distributions are homogeneous axially. Figure 8 now shows that the range of densities and temperatures spanned by all grid cell confined in the same volume as Figure 5 is much smaller. It is clear that the gas-prefill density is able to switch the plasma current rapidly, mitigating electro-thermal instabilities, while the possible MRT instabilities growing in the plasma sheet as it collapses do not imprint the rod before it gets compressed, limiting the source of non-uniformities.

At the time indicated in Figure 8, the maximum velocity of the sample is 400 m/s. At the given density, the density in kinetic energy

$$e_\text{K} = \frac{\rho v^2}{2} \tag{5}$$

is on the order of 0.2 J/mm³ while the pressure is close to 2kJ/mm³ (or 20 Mbar). Figure 8 is taken close to stagnation (where the radial velocity is close to 0) then we can say that the only two sources of energy at this time are the internal energy and the magnetic energy. At equilibrium (when v=0) we have the direct relationship

$$\nabla P = \mathbf{J} \mathrm{x} \mathbf{B}. \tag{6}$$

The radius of the warm dense matter sample at maximum compression was found to be 100 µm. This size is limited by our ability to capture the properties of warm dense matter at these pressure and by the resolution which is 50 µm. A more complete simulation, at higher resolution is required. However, if the equation of state of aluminum at such pressure is not fully known, increasing the resolution may not yield any information on the radius at full compression. As mentioned earlier, we see this paper as a proof-of-principle for an experimental campaign rather than demonstration that warm dense matter sample can be generated with pulsed-power driver and how degenerate and coupled this matter is.

## Conclusions

We have shown that a gas prefill can be used as an effective plasma opening switch. While the present work confirms previously published results obtained from two-dimensional MHD simulations, three dimensional simulations reveals that magnetic Rayleigh-Taylor instabilities pose a real threat to the practicality of the proposed scheme. However, a balance can be found where the prefill density is low enough to minimize the impact of instabilities on the final compression of the material sample while still be an effective plasma switch. The plasma switch not only steepens the current rise but it has very little impact on where the current flows after the switch has happened providing that the density of the gas prefill is low to limit the impact of magnetic Rayleigh Taylor instabilities. Numerical simulations find a factor of 5 increase in current rise with the right initial gas density.

However three-dimensional simulations where limited in resolution not allowing to verify the radial homogeneity of the density and temperature distribution profiles. However, we can expect the magnetic force to be effective only in the skin depth (depth of field penetration) which 10 µm. This number overestimates the skin depth since we assumed the material to be cold and not at 1 eV. This means that 90 % of the rod diameter is free of field at maximum compression. Since there is no JxB forces beyond that depth there should be also no pressure gradient at equilibrium (i.e. stagnation) as shown by Eq. (6). While the density and temperature profiles might not be flat, the pressure profile should be near constant since the sample.

There are two advantages to use a puled-power driver to produce warm dense matter. The first advantage is the cost of the driver which is low compare to usual drivers like high power lasers and heavy ion beams. The second advantage is the ability to confine warm dense matter for long periods of time (>1 ns) leaving enough time to measure equilibrium transport coefficients. The third advantage is the ability to generate large geometrical sample size (>100µm) allowing experimentalists to measure properties such as viscosity (or pressure) on scales as large as material grain size. When this size is invoked, the definition of viscosity (or pressure) is closer to the usual (macroscopic) interpretation rather than the less accepted microscopic definition.

**Acknowledgements**: This research was supported by the DOE grant number DE-SC0016252.

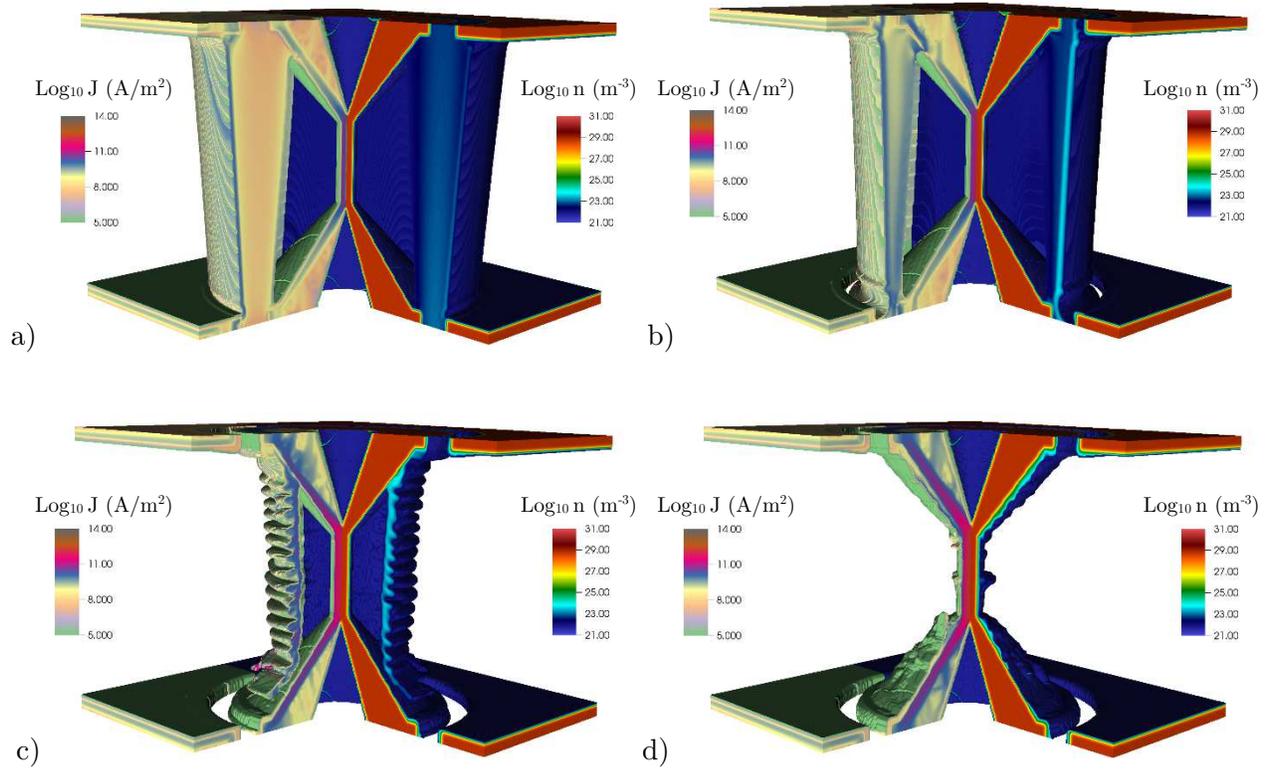

Figure 1. Time evolution of the current density (left) and ion number density (right) on the log scale for an initial gas prefill density of $16 \times 10^{22}$ m$^{-3}$ at a) t= 20ns, b) t=40 ns, c) t=60 ns and d) t= 80 ns. Volume information with density below $10^{22}$m$^{-3}$ is not printed.

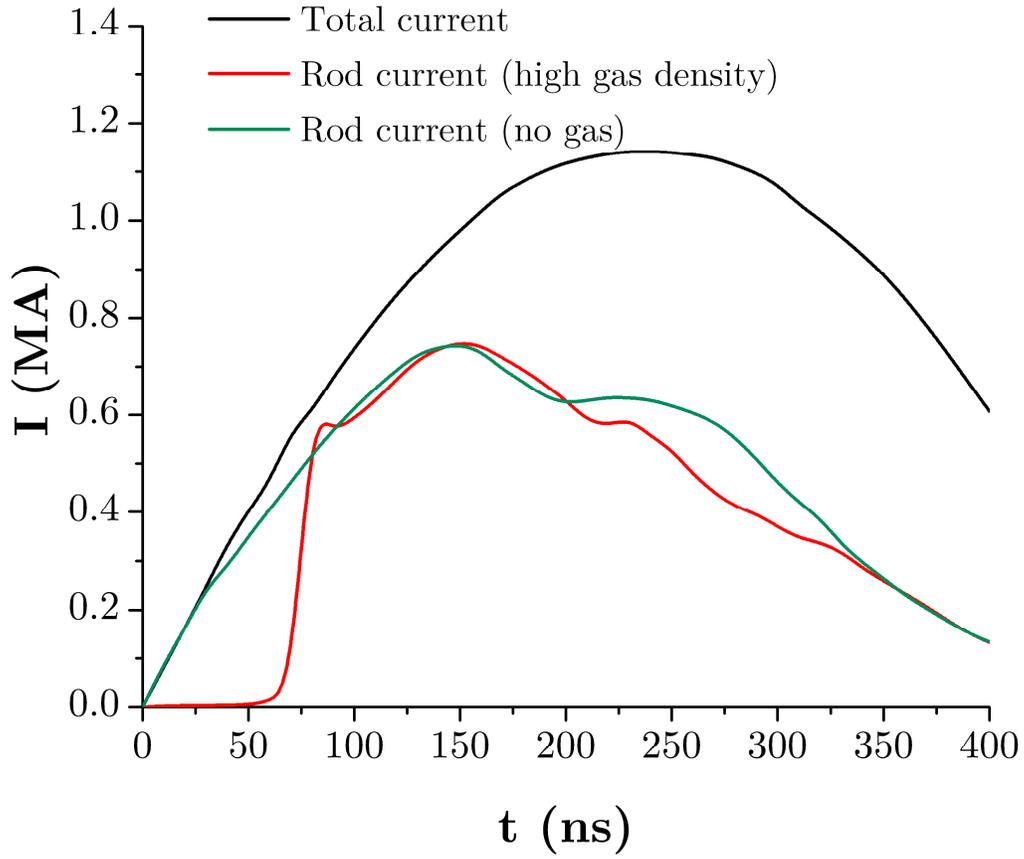

**Figure 2.** Time evolution of the total current discharge current flowing through the whole computational domain (black) plotted together with the rod currents flowing for an initial gas density of $16 \times 10^{22}$ m$^{-3}$ (red) and with no gas (green).

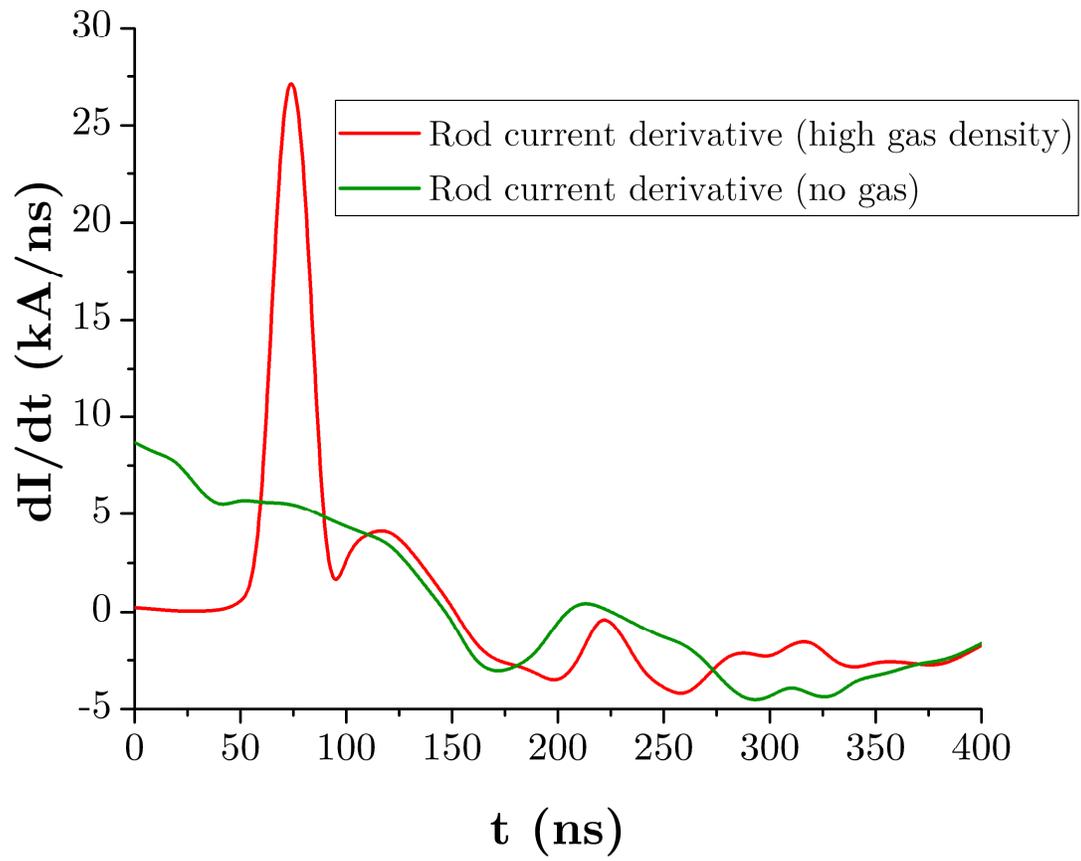

Figure 3. Time evolution of the rod current time derivative for gas density of $16 \times 10^{22}$ m$^{-3}$ (red) and with no gas (green)
13

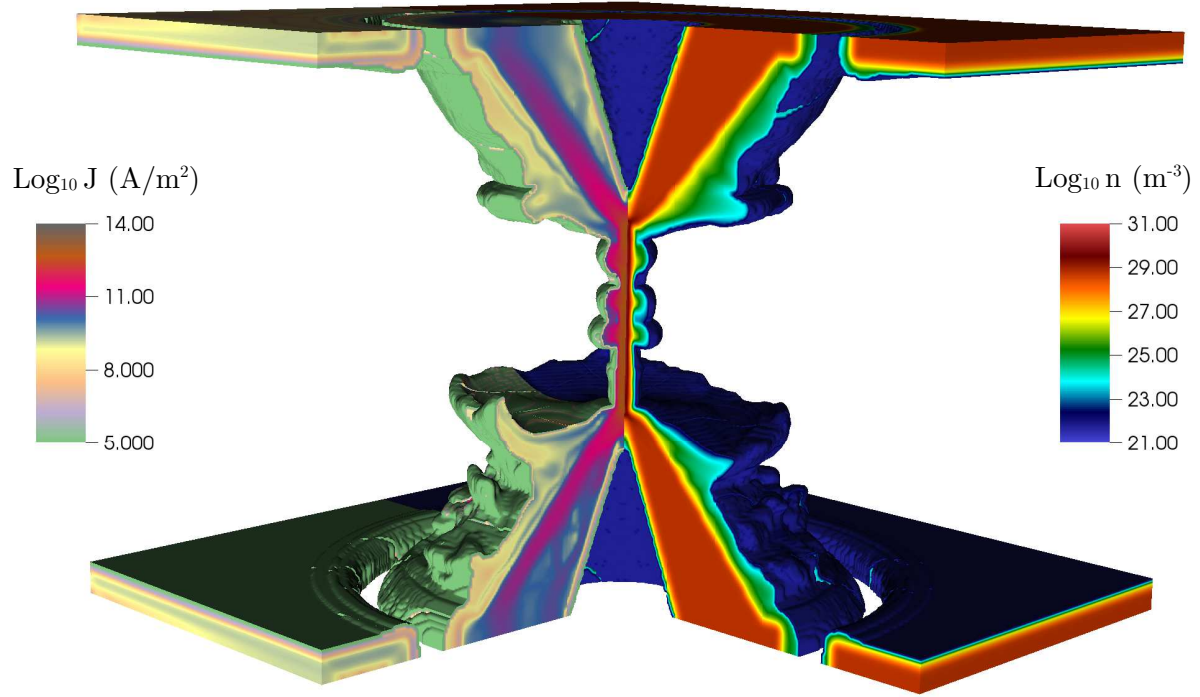

**Figure 4.** Current density (left) and plasma density (right) on the log scale at stagnation for an initial gas density of $16 \times 10^{22}$ m$^{-3}$ for t =165 ns.

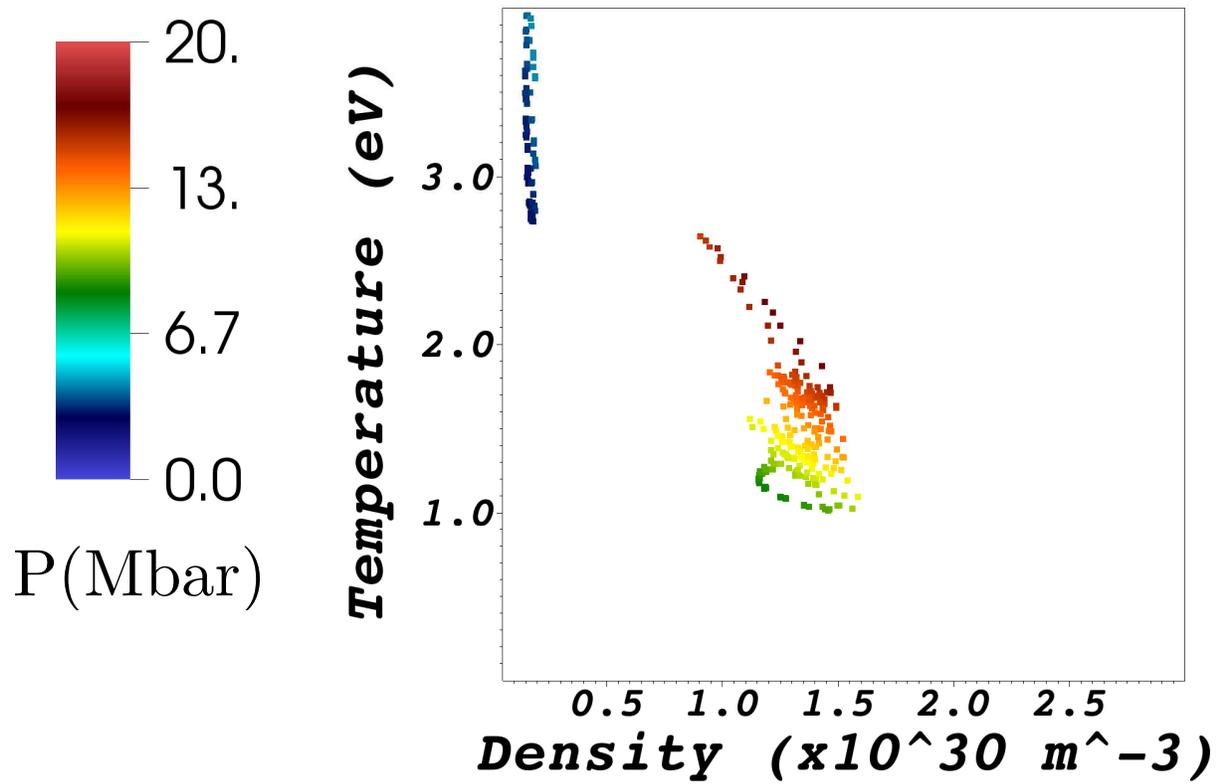

Figure 5. Ion number density versus the temperature for an initial gas density of $16 \times 10^{22}$ m$^{-3}$ at t=160 ns.

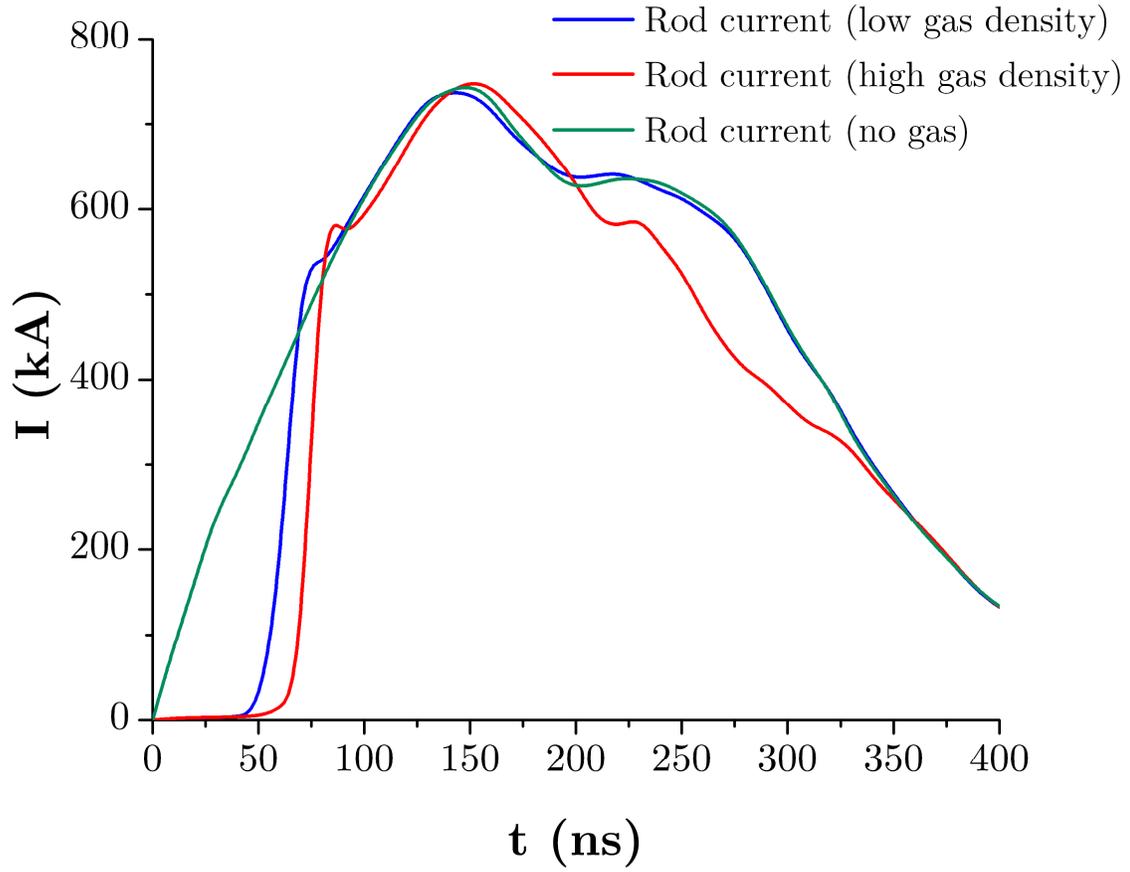

Figure 6. Time evolution of currents flowing through the rod for an initial gas density of $16\times10^{22}$ m$^{-3}$ (red), $8\times10^{22}$ m$^{-3}$ (blue) and 0 (green)

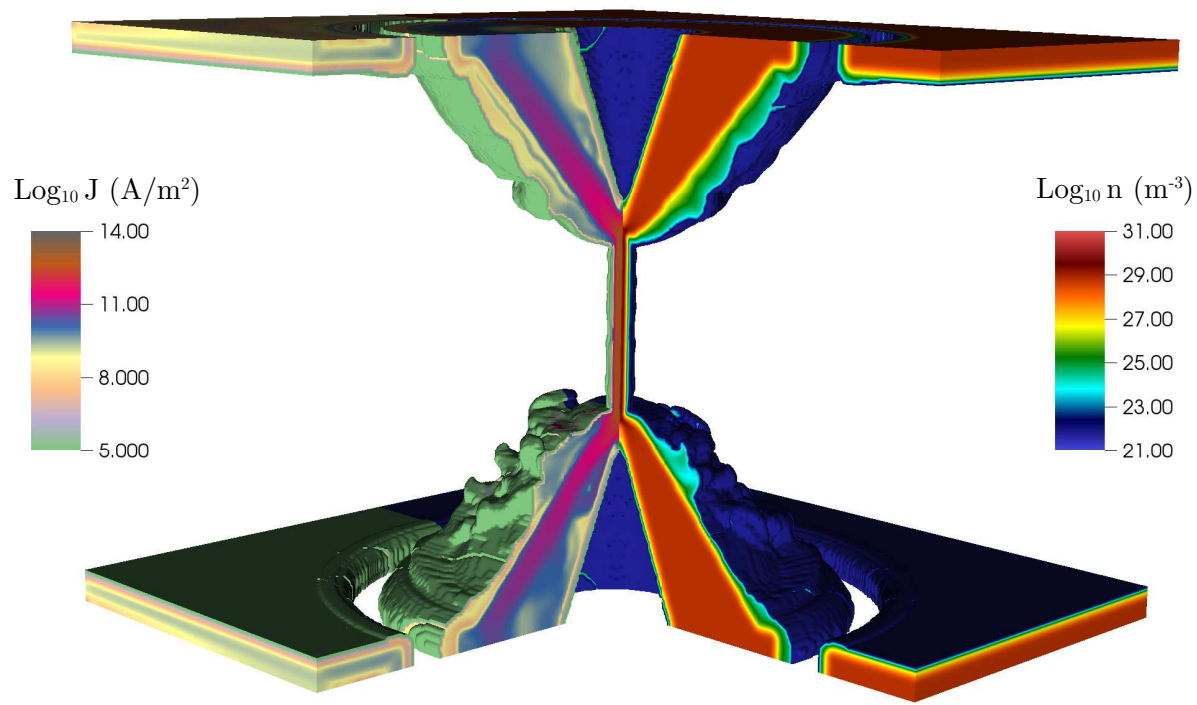

**Figure 7.** Current density (left) and plasma density on the log scale (right) at stagnation for an initial gas density of $8\times10^{22}$ m$^{-3}$ at t=150 ns.

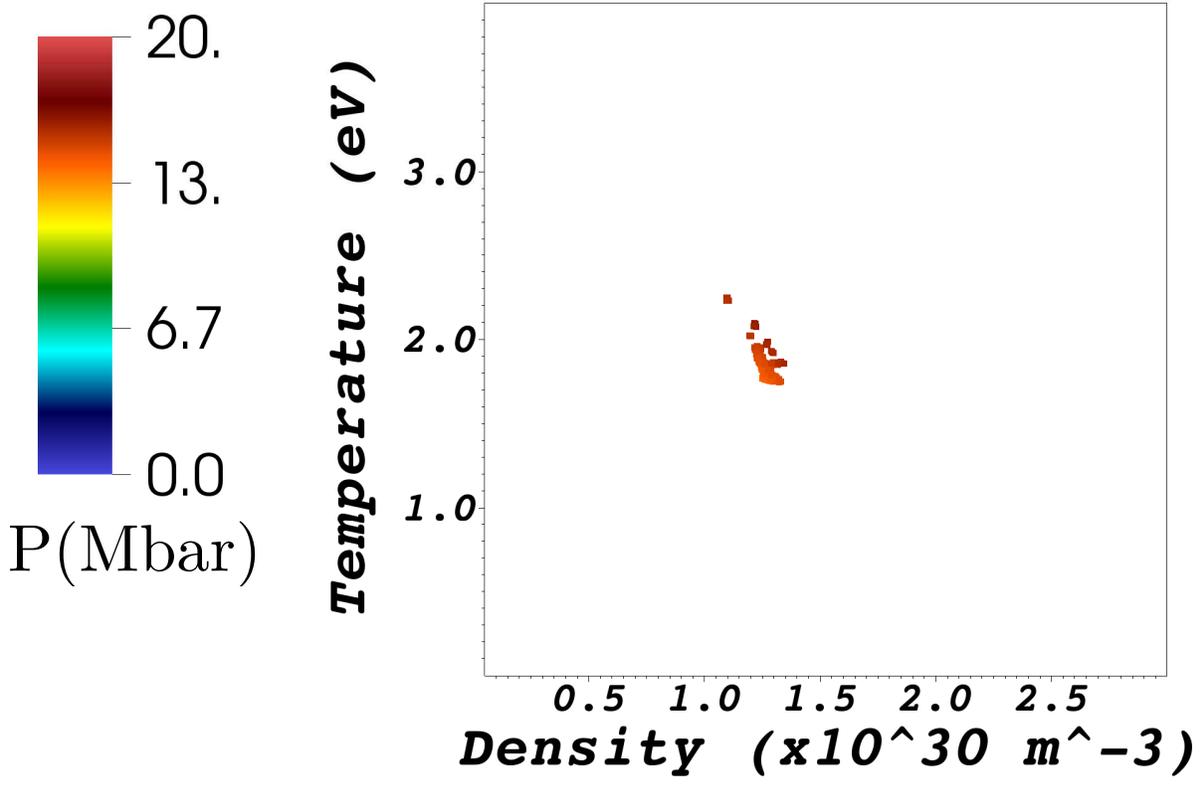

**Figure 8.** Ion number density versus the temperature for an initial gas density of 8x10²² m⁻³ at t=150 ns.